\documentclass[twocolumn,prb]{revtex4}
\usepackage{bm}
\usepackage{epsfig}

\begin{document}
\title{Spin Splitting  in Symmetrical SiGe Quantum Wells}

\author{L.E.~Golub}\email{golub@coherent.ioffe.ru}
\author{E.L.~Ivchenko}

\affiliation{A.F.~Ioffe Physico-Technical Institute, Russian Academy of Sciences, 194021 St.~Petersburg, Russia}

\date{\today}

\begin{abstract}
Spin splitting of conduction electron states has been analyzed for
all possible point symmetries of SiGe quantum well structures. A
particular attention is paid to removal of spin degeneracy caused
by the rotoinversion asymmetry of a (001) heterointerface between
two diamond-lattice materials. The asymmetry is shown to result in
spin splitting of both Rashba and Dresselhaus types in symmetrical
SiGe quantum wells. Consequences of the spin splitting on spin
relaxation are discussed.
\end{abstract}

\maketitle

Spin properties attract the great attention in recent years due to
attempts to realize an electronic device based on the spin of
carriers. Conduction electrons are obvious candidates for such
devices, particularly in nanostructures where electron energy
spectrum and shape of the envelope functions can be effectively
engineered by the growth design, application of electric or
magnetic fields as well as by illumination with light.

Various semiconductor materials are being involved in the
spintronics activities. SiGe quantum well (QW) structures are
among them.~\cite{ganichev,jantsch,jantsch_new,sherman} Although
bulk Si and Ge have an inversion center, QW structures grown from
these materials can lack such a center and allow the spin
splitting of the electronic subbands.

The quantum engineering of spintronic devices is usually focused
on the Rashba spin-dependent term to the electron effective
Hamiltonian in heterostructures. This contribution appears due to
asymmetry of the heteropotential (the so-called Structure
Inversion Asymmetry, or SIA) and has no relation to the properties
of a bulk semiconductor.

In III-V heterostructures, there exists another spin-dependent
contribution called the Dresselhaus term that appears due to Bulk
Inversion Asymmetry (BIA). It is commonly believed that the
Dresselhaus contribution is absent in structures grown from
centrosymmetric materials.

In the present work we show that the Dresselhaus-like spin
splitting  is possible in heterostructures made of Si and Ge due
to the anisotropy of chemical bonds at interfaces.

Symmetry of a (001)-grown interface between Si$_{1-x}$Ge$_x$ and
Si can be C$_{\rm 2v}$ or $C_{\rm 4v}$ on average.~\cite{xiao} The
former point group describes the symmetry of an ideal
heterointerface with the interfacial chemical bonds lying in the
same plane. A nonideal interface containing monoatomic
fluctuations has two kinds of flat areas with interfacial planes
shifted with respect to each other by a quarter of the lattice
constant. The local symmetry of each area is C$_{\rm 2v}$ as well.
However if the both kinds are equally distributed, the interface
overall symmetry increases up to $C_{\rm 4v}$. It follows then
that the symmetry of a Si$_{1-x}$Ge$_x$/Si QW structure containing
two interfaces is described by one of five point groups: $D_{\rm
2d}$ or $D_{\rm 2h}$ in case of two ideal interfaces with odd or
even number of monolayers between them; $C_{\rm 2v}$ for a pair of
ideal and rough interfaces; $C_{\rm 4v}$ or $D_{\rm 4h}$ for two
non-ideal interfaces of the overall symmetry $C_{\rm 4v}$ each,
see Fig.~1.

\begin{figure*}
\epsfxsize=\textwidth \epsfysize=0.2\textwidth
\centering{\epsfbox{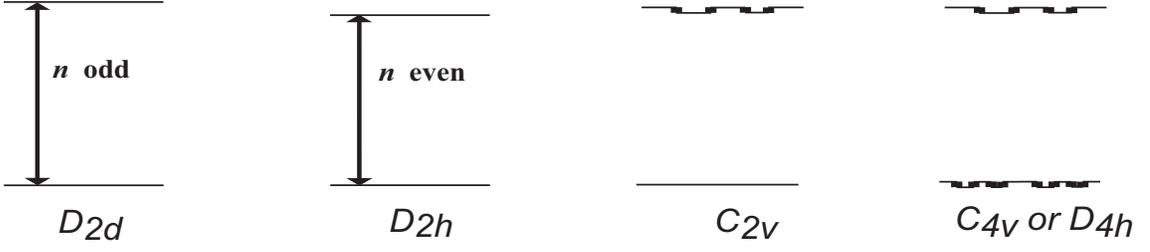}} \caption{Different interface
profiles and QW point symmetries. The growth direction is [001],
$n$ is a number of monoatomic layers.}
\end{figure*}

Two of the above-mentioned groups, $D_{\rm 4h}$ and $D_{\rm 2h}$,
contain the space inversion operation and forbid the spin
splitting of electronic states. Three remaining groups allow the
spin-dependent linear-in-${\bm k}$ term ${\cal H}_{{\bm k}} =
\gamma_{\alpha \beta} \sigma_{\alpha} k_{\beta}$. Here
$\sigma_{\alpha}$ are the Pauli matrices and ${\bm k} = (k_x,
k_y)$ is the in-plane electron wave vector.

In Si$_{1-x}$Ge$_x$/Si QWs grown along direction $z \parallel
[001]$ with low enough content of Ge in the alloy layer, the
lowest conduction band is located near the $X$ point of the
Brillouin zone. Note that in the following we consider the
electronic states attached to the $X_z$ valley because, due to the
quantum confinement effect, its bottom lies lower than $X_x$ and
$X_y$ valleys.~\cite{aleshkin} Hereafter the subscripts $x,y,z$
indicate three $X$ valleys as well as the Cartesian coordinate
system with $x \parallel [100]$, $y \parallel [010]$ being the
in-plane axes. The symmetry analysis of the electron quantum
confined states is based on the fact that at the $X$ point the
bulk Bloch functions form projective representations of the point
group $D_{\rm 4h}$. All five above-mentioned groups are subgroups
of $D_{\rm 4h}$. In the group $D_{\rm 2d}$, the Dresselhaus-like
term
\[
h_D({\bm k}) = \sigma_x k_x -\sigma_y k_y
\]
is an only invariant which can be constructed from the products
$\sigma_{\alpha} k_{\beta}$. On the contrary, in the group $C_{\rm
4v}$, the only invariant combination is the Rashba term
\[h_R({\bm k}) = \sigma_x k_y - \sigma_y k_x.\]
The analysis shows that both combinations $h_D({\bm k})$ and
$h_R({\bm k})$ are invariants of the group $C_{\rm 2v}$, i.e. both
Dresselhaus- and Rashba-like spin-dependent terms are allowed.

In SiGe/Si QWs with high Ge content, the lowest conduction band
may be located at the $L$ point. The bulk Bloch functions form
projective representations of the $D_{\rm 3d}$ point group which
are $p$-equivalent to the usual representations of the same group.
In this case the associated coordinate system $x',y',z'$ is
connected with the principal valley axis $z' \parallel [111]$ and
the in-plane axes $x' \parallel [1 \bar{1} 0]$ (perpendicular to
one of the mirror-reflection planes) and $y' \parallel [1 1
\bar{2}]$. In (001)-grown QWs of the symmetry $D_{\rm 2h}$ or
$D_{\rm 4h}$, intersection of these point groups with $D_{\rm 3d}$
is $C_{\rm 2h}$ with the inversion, so that the spin degeneracy is
retained. On the other hand, the intersection of the rest three
groups with $D_{\rm 3d}$ is $C_{\rm s}$. As a result, the
effective Hamiltonian for the $L$-valley electrons in QWs of the
symmetry $D_{\rm 2d}$, $C_{\rm 2v}$ or $C_{\rm 4v}$ contains three
linearly independent combinations $\sigma_{y'} k_{x'}$,
$\sigma_{z'} k_{x'}$ and $\sigma_{x'} k_{y'}$ responsible for the
spin splitting.

SiGe/Si heterostructures grown in the (111)-direction have either
$D_{\rm 3d}$ or $C_{\rm 3v}$ point symmetry depending on the
parity of monolayer numbers. The $D_{\rm 3d}$ group has the
inversion center and retains the spin degeneracy whereas the
$C_{\rm 3v}$ group allows the spin splitting. The lowest
conduction band in (111)-SiGe/Si QWs is located at the $L$ point
of the Brillouin zone. Since $C_{\rm 3v}$ is a subgroup of $D_{\rm
3d}$, the relevant symmetry is $C_{\rm 3v}$, and the invariant
combination of the products $\sigma_{\alpha} k_{\beta}$ is the
Rashba term $$\sigma_{x'} k_{y'} - \sigma_{y'} k_{x'}.$$

Microscopically, a linear-in-$\bm k$ correction to the
conduction-band effective Hamiltonian is given by the second order
of the perturbation theory
\begin{equation}\label{second_order}
\Delta {\cal H} = { H_{\rm SO} H_{\bm {k p}}^\dag + H_{\bm {k p}}
H_{\rm SO}^\dag
 \over E_c - E_v}.
\end{equation}
Here $H_{\bm{k p}}$ and $H_{\rm SO}$ are interband $\bm{k \cdot
p}$ and spin-orbit interaction Hamiltonians, respectively. We take
into consideration only the coupling of the conduction $X_1$- and
the valence $X_4$-states. The symmetry properties of the Bloch
functions at the $X$ point of a diamond-lattice semiconductor
crystal coincide with those of the following
functions~\cite{yu/cardona}
\begin{equation}
\label{X1} X_1\mbox{-states}: \left\{
\begin{array}{cc}
S=\cos{\left( {2 \pi z / a}\right)} \\
Z=\sin{\left( {2 \pi z / a}\right)}
\end{array}
\right.,
\end{equation}
\begin{equation}
\label{X4} X_4\mbox{-states}: \left\{
\begin{array}{cc}
X = \sin{\left( {2 \pi x / a}\right)} \cos{\left( {2 \pi y /
a}\right)} \\
Y = \cos{\left( {2 \pi x / a}\right)} \sin{\left( {2 \pi y /
a}\right)}
\end{array}
\right. ,
\end{equation}
where $a$ is the lattice constant. For the bulk states $X_1, X_4$ in the bases~(\ref{X1}),~(\ref{X4}), the interband matrix elements can be presented as
\begin{equation} \label{kpso}
H_{\bm{k p}} =  {\cal P} \left( \begin{array}{cc}
k_x & k_y\\
-k_y & -k_x \end{array} \right), \:\:\:\:
H_{\rm SO} = {\cal U} \left( \begin{array}{cc} \sigma_x & -\sigma_y\\
-\sigma_y & \sigma_x
\end{array}
\right).
\end{equation}
Here ${\cal P} = (\hbar/m_0) \left< S | \hat{p}_x | X \right>$,
${\cal U} = \left< S | U_x | X \right>$, $\hat{\bm p}$ is the
momentum operator, and the pseudovector ${\bm U} = (\hbar / 4
m_0^2 c^2) {\bm \nabla} W \times \hat{\bm p}$ enters into the
spin-orbit Hamiltonian $H_{\rm SO} = {\bm \sigma} \cdot {\bm U}$
($W$ is the microscopic potential).

The substitution of (\ref{kpso}) into (\ref{second_order}) results
in
\[
\Delta {\cal H} \propto h_D({\bm k}) \left( \begin{array}{cc} 1 & 0\\
 0 & -1 \end{array} \right),
\]
where the second multiplier is a 2$\times$2 matrix related to the
basis (\ref{X1}). The matrix $\Delta {\cal H}$ {\em does not} lead
to lifting the degeneracy of the $X_1$-states in the bulk
centrosymmetric material, in accordance with the general symmetry
consideration.

The splitting can be achieved if one takes into account the
anisotropy of chemical bonds at the interfaces. It results in
$\delta$-functional contributions to $H_{\bm{k p}}$ and $H_{\rm
SO}$ of the form
\begin{equation} \label{deltakpso}
\Delta H_{\bm{k p}} = V_{\bm{k p}}\: \delta(z - z_{if}),\:\:\:
\Delta H_{\rm SO} = V_{\rm SO}\: \delta(z - z_{if}),
\end{equation}
where $z_{if}$ is the interface coordinate, and the matrices
$V_{\bm{k p}}, V_{\rm SO}$ have few linearly independent
components. In the case of the lowest symmetry under study,
$C_{\rm 2v}$, each matrix is determined by four
linearly-independent parameters
\begin{eqnarray} \label{kp_C2v}
V_{\bm{k p}} &=& \left(
\begin{array}{cc}
P_1 k_x + P_2 k_y & P_1 k_y + P_2 k_x\\ P_3 k_y + P_4 k_x & P_3
k_x + P_4 k_y\end{array} \right), \\
 \label{so_C2v}
V_{\rm SO} &=& \left(
\begin{array}{cc}
U_1 \sigma_x - U_2 \sigma_y & -U_1 \sigma_y + U_2 \sigma_x\\ -U_3
\sigma_y + U_4 \sigma_x & U_3 \sigma_x - U_4 \sigma_y\end{array}
\right).
\end{eqnarray}
In the bases~(\ref{X1}),~(\ref{X4}),  the parameters $P_n$ and
$U_n$ $(n=1 \div  4)$ are purely imaginary.

By using Eqs.~(\ref{second_order}) and
(\ref{deltakpso})--(\ref{so_C2v}), one can show that, for a single
interface of the $C_{\rm 2v}$-symmetry, the correction to the
conduction-band Hamiltonian linear in $k_{x,y}$ and responsible
for the removal of spin degeneracy is given by
\begin{equation} \label{c2vif}
\Delta {\cal H}_{C_{\rm 2v}} = \left( \begin{array}{cc} 1 & 0\\
 0 & 1 \end{array} \right) H_{if} \delta(z - z_{if}) \:.
\end{equation}
Here $H_{if}$ is a linear combination of the spin Pauli matrices
that, in the first-order approximation in the
perturbations~(\ref{deltakpso}), has the form
\begin{eqnarray}
\label{H_C2v} H_{if} &=& {{\cal P} \over E_0} \biggl[(U_3 - U_1)
\: h_D({\bm k}) + (U_4 - U_2)\: h_R({\bm k}) \biggr]
\\
&-& {{\cal U} \over E_0} \biggl[ (P_1 + P_3) \: h_D({\bm k}) +
(P_2 + P_4) \: h_R({\bm k}) \biggr], \nonumber
\end{eqnarray}
with $E_0$ being the band gap between $X_1$ and $X_4$ states.

The electron effective Hamiltonian in an ideal QW contains a sum
of two contributions (\ref{c2vif}) related to the left-~($l$) and
right-hand-side ($r$) interfaces. If the QW contains an even
number of monoatomic layers ($D_{\rm 2h}$ symmetry), then the
corresponding parameters $U_n^l, U_n^r$ or $P_n^l, P_n^r$ are
interconnected by
\[
[U_1, U_2, U_3, U_4]_r = [U_3, U_4, U_1, U_2]_l, \]
\[
[P_1, P_2, P_3, P_4]_r = [- P_3, - P_4, - P_1, -P_2]_l,
\]
and if the QW contains an odd number of monoatomic layers ($D_{\rm
2d}$ symmetry), then
\[
[U_1, U_2, U_3, U_4]_r = [U_1, - U_2, U_3, - U_4]_l, \]
\[
[P_1, P_2, P_3, P_4]_r = [P_1, -P_2, P_3, -P_4]_l.
\]
From here it readily follows that, in the former case, the
electronic states in SiGe QWs are spin-degenerate ($H_{if}=0$)
and, in the latter, the spin degeneracy is removed by a
Dresselhaus-like linear-in-${\bm k}$ terms. The similar analysis
can be carried out for non-ideal SiGe QW structures of the $C_{\rm
2v}$ and $C_{\rm 4v}$ symmetry.

\emph{Spin splitting of $L$-electrons}. Without spin, the basis
functions for the $L$-point may be chosen as $s$-like for the
$L_1$ conduction-band state (the Bloch function $S'$), and as
$p$-like ($X',Y'$) for the $L_{3'}$ valence-band states. With
spin, they are multiplied by the spin functions $\uparrow$ and
$\downarrow$.

In a bulk diamond-lattice semiconductor, the spin-orbit
interaction results in a splitting by some value $\Delta$ of the
$L_{3'}$ valence band into the $(L_4^-,L^-_5)$ and $L_6^-$
subbands. In the three-fold degenerate $\Gamma$-point, the
second-order perturbation theory with allowance for $\Delta \neq
0$ gives rise to Dresselhaus-like conduction-band spin-splitting
in III-V QWs.~\cite{Roessler/Kainz} However, this is not the case
for the $L$-point due to large energy spacing between the top
valence band $L_{3'}$ ($X', Y'$) and lower lying valence-band
states. Therefore, one should again take into account the
interband spin-orbit interaction.

In the bulk, the spin-orbit coupling between the $L_1$ and
$L_{3'}$ bands is forbidden. However, the group $C_{\rm s}$ has
only two symmetry operations: identical and reflection in the
$(y',z')$-plane. The reflection yields $\left< S' | U_{x'} | X'
\right> = \left< S' | U_{y',z'} | Y' \right> = 0$, but the
following matrix elements are nonzero:
\begin{eqnarray}
\label{U_L} U'_1 &=& \left< S' | U_{y'} | X' \right>, \nonumber\\
U'_2 &=& \left< S' | U_{z'} | X' \right>,\\ U'_3 &=& \left< S' |
U_{x'} | Y' \right>. \nonumber
\end{eqnarray}

Introducing the matrix element of the bulk $\bm{k \cdot
p}$-interaction $\mathcal{P}' = (\hbar / m_0) \left< S' |
\hat{p}_x | X' \right>$, we obtain from Eqs.~(\ref{second_order})
and~(\ref{U_L}) for the $L$-states in a (001)~QW
\begin{equation}\label{H_Cs}
    \Delta {\cal H}_{C_{\rm S}} = - {2 \mathcal{P}' \over E'_0}
\delta(z-z_{if}) (U'_2 \sigma_{z'} k_{x'} + U'_1 \sigma_{y'}
k_{x'} + U'_3 \sigma_{x'}  \langle k_{y'} \rangle).
\end{equation}
Here $E'_0$ is the energy gap between the $L_1$ and the $L_{3'}$
states and  the angular brackets mean averaging over the
quantum-confined state. As a result the component $k_{y'} = (k_x +
k_y - 2 k_z)/\sqrt{6}$ reduces to $\langle k_{y'} \rangle = (k_x +
k_y)/\sqrt{6}$ whereas the component $k_{x'} = (k_x -
k_y)/\sqrt{2}$ remains unchanged.

Inclusion of the spin splitting $\Delta$ of the valence band
slightly modifies the coefficients in the above expression:
\begin{eqnarray}
  \Delta {\cal H}_{C_{\rm S}} &=& - \delta(z-z_{if}) \mathcal{P}'  \nonumber\\
  &\times& \left[
  \left( {1 \over E'_0 - \Delta/2} + {1 \over E'_0 + \Delta/2} \right) U'_2 \sigma_{z'} k_{x'}
  \right. \nonumber \\
&+&  \left. \left( {U'_1 + U'_3 \over E'_0 - \Delta/2} + {U'_3 -
U'_1 \over E'_0 + \Delta/2} \right) \sigma_{x'}  \langle k_{y'}
\rangle
  \right.  \nonumber \\
  &+& \left.
\left( {U'_1 + U'_3 \over E'_0 - \Delta/2} + {U'_1 - U'_3 \over
E'_0 + \Delta/2} \right) \sigma_{y'} k_{x'} \right]. \nonumber
\end{eqnarray}

In (111)-grown QWs the relevant symmetry is $C_{\rm 3v}$ with the
axis $C_3$ and two more mirror reflection planes in addition to
the elements of $C_{\rm s}$ group. The rotation yields
\[
    U'_1 = - U'_3, \:\:\:\:\:\: U'_2=0,
\]
and we get from Eq.~(\ref{H_Cs}) the Rashba-like contribution for
the (111) SiGe QWs with odd number of monolayers:
\begin{equation}\label{H_C3v}
    \Delta {\cal H}_{C_{3 \rm V}} = \delta(z'-z_{if}) {2 \mathcal{P}' U'_1 \over
    E'_0} \: (\sigma_{x'} k_{y'} - \sigma_{y'} k_{x'}).
\end{equation}
If the number of monolayers is even ($D_{3 \rm d}$ symmetry) then
the inversion imposes the condition $U'_1 = 0$, and spin splitting
is absent.

Within the envelope function approximation, the electron wave
function $\psi$ satisfies the Schr\"odinger equation with the
effective Hamiltonian $${\cal H} = {\cal H}_0(k_z, {\bm k}) + V(z)
+ \Delta {\cal H}.$$ Here ${\cal H}_0$ is the bulk
spin-independent Hamiltonian with $k_z = - \mbox{i} \partial /
\partial z$, $V(z)$ is the heteropotential, and the correction
$\Delta {\cal H}$ is given by Eq.~(\ref{c2vif}).

Instead of solving the Schr\"odinger equation with the Hamiltonian
${\cal H}$, one can equivalently find general solutions of the
equations $({\cal H}_0 + V) \psi = E \psi$ within each homogeneous
layer and then apply the boundary conditions
\begin{eqnarray}\label{BCond}
\psi(z_{if}-0) &=& \psi(z_{if}+0), \\
\left. (v_z \psi) \right|_{z_{if}-0} &=& \left. (v_z \psi)
\right|_{z_{if}+0} + {2 \mbox{i} \over \hbar} H_{if} \:
\psi(z_{if}), \nonumber
\end{eqnarray}
where the velocity operator $v_z = \hbar^{-1} \partial {\cal
H}_0/\partial k_z$.

In order to estimate the spin splitting one needs to go beyond the
envelope function approximation. This can be done in the
pseudo-potential or tight-binding model which yields the matrix
$H_{if}$ in boundary conditions~(\ref{BCond}). The work on
estimation of the interface-induced spin splitting in the
microscopic tight-binding model is in progress. Here it suffices
to note that the spin matrix $H_{if}$ in Eq.~(\ref{H_C2v}) is
non-vanishing if interatomic contributions to $H_{\rm SO}$ either
$H_{\bm {kp}}$ are taken into account.

In this paper, we have analyzed spin splitting of electron states
in SiGe heterostructures of all possible symmetries. The absence
of inversion center can be also probed by means of second-harmonic
generation experiments.~\cite{xiao,bottomley,SHG2} An ideal
Si$_m$Ge$_n$ superlattice with odd $n$ and $m$ allows
second-harmonic generation. It was demonstrated experimentally
that Si$_m$Ge$_n$ superlattices with nominally odd and even $n$,
$m$ possess comparable second-harmonic conversion
efficiency.~\cite{bottomley} The weak nonlinear response can be
explained by antiphase microscopic domains shifted with respect to
each other by one monoatomic layer along the growth direction
[001].

Similarly, different domains are characterized by opposite signs
of the linear in ${\bm k}$ spin-dependent matrix $H_{if}$ in
Eq.~(\ref{c2vif}). The influence of this kind of imperfection on
the D'yakonov-Perel' spin relaxation of the conduction electrons
depends on the relation between the linear dimension, $l_D$, of a
single domain and the electron mean free path, $l$, in the
interface plane. In the case $l_D \gg l$, the spin relaxation time
of free carriers is the same as in a perfect structure:
\begin{equation}\label{taus}
\frac{1}{\tau_s} = \frac{2}{\hbar^2} \left< \tau_p \mbox{Tr}
\left( H_{if}^2 \right) \right> \left[ \psi^\dag(z_{if})
\psi(z_{if}) \right]^2  \:.
\end{equation}
Here $\tau_s$ is the relaxation time of spin $z$-component (for
in-plane spin the corresponding time is $2 \tau_s$). In
Eq.~(\ref{taus}) $\tau_p$ is the momentum scattering time, the
angular brackets mean averaging over the carrier energy
distribution, and $\psi(z_{if})$ is the interface value of the
envelope function calculated at $H_{if}=0$. In the opposite
limiting case $l_D \ll l$, one has
\begin{equation} \label{tausd}
\frac{1}{\tau_s(l_D)} \approx \frac{1}{\tau_s} \: \frac{l_D}{l}\:,
\end{equation}
where $\tau_s$ is given by Eq.~(\ref{taus}). If the time $\tau_p$
is governed by scattering from the antiphase-domain walls
(boundaries) one can use Eq.~(\ref{taus}) for estimations of the
spin relaxation times. Thus, we conclude that, even if the overall
symmetry of a SiGe heterostructure is $D_{4 \rm h}$ due to the
antiphase domains, the lack of inversion center within a
particular domain leads to the spin dephasing according to
Eqs.~(\ref{taus}) and (\ref{tausd}). This opens a possibility to
discuss the recent spin relaxation times measurements in these
heterostructures.~\cite{jantsch,jantsch_new}

In conclusion, we have shown that linear-in-${\bm k}$ spin
splitting is present even in symmetrical SiGe QWs. It can be of
Rashba, Dresselhaus, or both types. The splitting is caused by
anisotropy of chemical bonds at interfaces.

\mbox{}\\

This work is financially supported by
the RFBR, INTAS, ``Dynasty'' Foundation --- ICFPM, and by the
Programmes of RAS and Russian Ministries of Science and Education.

\end{document}